\documentclass[12pt]{article}
\usepackage{amsmath}
\usepackage[dvips]{graphicx}
\usepackage{epsfig}
\usepackage{amsmath}
\usepackage{amssymb}
\usepackage{graphics,epstopdf}
\usepackage{amsfonts}
\usepackage{amsthm}
\usepackage[usenames]{color}

\definecolor{AV}{rgb}{0.65,0.0,0}
\definecolor{GC}{rgb}{0,0.0,0.65}
\definecolor{WS}{rgb}{0,0.65,0}

\usepackage[
      colorlinks=true,
      linkcolor=blue,
      urlcolor=blue,
      filecolor=blue,
      citecolor=red,
      pdfstartview=FitV,
      pdftitle={},
      pdfauthor={},
      pdfsubject={},
      pdfkeywords={},
      pdfpagemode=None,
      bookmarksopen=true
]{hyperref}

\newcommand{\bm}{\begin{multiline}}
\newcommand{\beq}{\begin{equation}}
\newcommand{\eeq}{\end{equation}}
\newcommand{\beqs}{\begin{eqnarray}}
\newcommand{\eeqs}{\end{eqnarray}}

\begin{document}

\thispagestyle{empty}

\hfill{}

\hfill{}

\hfill{}

\vspace{32pt}

\begin{center}

\textbf{\Large Massless fermions on static general prolate metrics and their Heun solutions}

\vspace{48pt}

\textbf{Marina-Aura Dariescu,}\footnote{E-mail: \texttt{marina@uaic.ro}}
\textbf{Ciprian Dariescu, }\footnote{E-mail: \texttt{ciprian.dariescu@uaic.ro}}
\textbf{Cristian Stelea,}\footnote{E-mail: \texttt{cristian.stelea@uaic.ro}}

\vspace*{0.2cm}

\textit{$^{1,2}$ Faculty of Physics, ``Alexandru Ioan Cuza" University}\\[0pt]
\textit{11 Bd. Carol I, Iasi, 700506, Romania}\\[.5em]

\textit{$^3$ Research Department, Faculty of Physics, ``Alexandru Ioan Cuza" University}\\[0pt]
\textit{11 Bd. Carol I, Iasi, 700506, Romania}\\[.5em]

\end{center}

\vspace{30pt}

\begin{abstract}
Employing a pseudo-orthonormal coordinate-free approach we write down the Dirac equation in spacetimes with static generally prolate metrics. As examples, we consider the electrically charged C-metric, the vacuum C-metric, the Reissner-Nordstrom and Schwarszchild spacetimes and the BBMB black hole and show that the solutions to the Dirac equations for particles these spacetimes can be derived in terms of Heun's general functions and their confluent and double confluent forms.   
\end{abstract}

\begin{flushleft}
{\it Keywords}: Dirac Equation; Heun functions; C-metric.
\\ 
{\it PACS:}
04.20.Jb ; 02.40.Ky ; 04.62.+v ; 02.30. Gp; 11.27.+d.
\end{flushleft}

\baselineskip 1.5em

\newpage

\section{Introduction}

Even though the C-metric has a long history, being discovered by Levi-Civita one hundred years ago \cite{LC}
and later re-obtained by Ehlers and Kundt \cite{EK}, this nonlinear superposition of the Schwarzschild black hole solution and the Rindler flat spacetime associated with uniformly accelerated observers \cite{Griffiths:2006tk} (for a comprehensive review of its properties see \cite{grifpod}) can be seen as an unique laboratory to test the properties of the fields evolving in an accelerating background.

On the other hand, the study of the Dirac equation on different types of black holes is a subject with a long-standing history, the behavior of fermions in these backgrounds has been tackled mainly with approximate methods or numerical techniques \cite{Huang:2017nho}. Recently, the problem of the quantum fermions scattered from Schwarzschild black holes was studied using a version of partial wave analysis that allowed the authors to write down closed formulas for the scattering amplitudes and cross sections \cite{Cotaescu:2016aty}. However, similarly to the analysis performed in the uncharged Schwarzschild case, in the presence of a spherically symmetric static electromagnetic field, the radial part of the massless Dirac equation has been put into a Heun-type form \cite{Birkandan:2017rdp}, \cite{Hortacsu:2011rr}.

In terms of the techniquesused in these studies, the Newman-Penrose (NP) formalism \cite{Newman:1961qr} is widely considered to be a valuable tool for decoupling and separation of variables when dealing with the Dirac equation describing fermions in the vicinity of
different types of black holes \cite{He:2006jv}.

For the vacuum C-metric spacetime, in \cite{Bini:2015xpa},
by defining a Kinnersley-like null frame \cite{Kinnersley:1969}, the standard NP formalism is employed to solve the Dirac equation for massless particles. The exact solutions for the radial and angular
equations have been derived in terms of Heun functions  \cite{Ronveaux}, \cite{Slavyanov}, which generalize well-known special functions such as the Spheroidal Wave, the Lame or the hypergeometric functions.

As an alternative approach to the one mentioned above, in the present paper, we are developing a free of coordinates method, based on Cartan's formalism \cite{Dariescu:2017ima}. Thus, we are computing all the geometrical essentials for writing down the Dirac equation in its $SO(3,1)-$gauge covariant formulation, generalizing existent studies, as for example \cite{Casals:2012es}, where the authors are switching between canonical and pseudo-orthonormal basis and the solutions are derived using numerical techniques. As particular cases, besides the vacuum C-metric, in the present paper, a special attention is given to the Reissner-Nordstrom spacetime as well as to the BBMB metric, describing the black hole found by Bocharova, Bronnikov and Melnikov \cite{Bocharova:1970skc} and studied later by Bekenstein \cite{Bekenstein:1974sf}. 

The structure of this work is as follows: in the next section using the orthonormal Cartan formalism we derive the general form of the Dirac equation for a general metric  with axial symmetry. In the particular case of the metric (\ref{metric2}) we show that the massless Dirac equation can be reduced to two systems of decoupled equations (\ref{dec1}) and (\ref{dec2}) for the radial and angular functions. In section $3$ we show that in the background of the charged C-metric given in \cite{Hong:2003gx} one can express the general solution of the massless Dirac equation in terms of the general Heun functions. In section $4$ we show how to recover previous results known in literature in the particular cases of the uncharged C-metric, the Reissner-Nordstrom metric and the BBmB-metric. The last section is dedicated to conclusions and avenues for further work.

\section{The $SO(3,1)-$gauge covariant Dirac Equation}

Let us start with the general static prolate metric of the form \cite{Dariescu:2017ima}
\begin{equation}
ds^2=e^{2f} dr^2 + a^2 e^{2p} d \theta ^2 + b^2 e^{2q} d \varphi^2 - e^{2h} dt^2,
\label{generalm}
\end{equation}
where the functions $f$, $p$, $q$ and $h$ depend only on the coordinates $r$ and $\theta$.

Within the $SO(3,1)-$gauge covariant formulation, we introduce the pseudo-orthonormal frame $\lbrace E_a \rbrace_{(a=\overline{1,4})}$, i.e.
\[
E_1= e^{-f} \partial_r  \; , \; \;
E_2= \frac{e^{-p}}{a}  \, \partial_{\theta} \; , 
\; \; E_3 = \frac{e^{-q}}{b}  \, \partial_{\varphi} \; , \; \;
E_4 = e^{-h} \, \partial_t \; ,  
\]
whose corresponding dual base is 
\[
\omega^1= e^f \, dr\; , \; \;
\omega^2= a e^p \,d\theta \; , 
\; \; \omega^3= b e^q \, d\varphi \; , \; \;
\omega^4= e^h \,dt \; ,  
\]
so that the metric (\ref{generalm}) becomes $ds^2 = \eta_{ab} \, \omega^a \omega^b$, where $\eta_{ab} = diag \left[ 1 , \, 1 , \, 1 , \, -1 \right]$ is the usual Minkowsky metric. 

Using the first Cartan's equation, 
\begin{eqnarray}
d\omega^a&=&\Gamma^a_{.[bc]}\,\omega^b\wedge \omega^c \, ,
\end{eqnarray}
with $1 \leq b<c \leq4$ and $\Gamma^a_{.[bc]} = \Gamma^a_{.bc} - \Gamma^a_{. cb}$,
we obtain the following connection one-forms $\Gamma_{ab} = \Gamma_{abc} \omega^c$, where $\Gamma_{abc} = - \Gamma_{bac}$ , namely
\begin{eqnarray}
\Gamma_{12} &= & f_{|2}\,\omega^1 -  p_{|1}\,\omega^2 = 
\frac{e^{-p}}{a} \frac{\partial f}{\partial \theta} \, \omega^1 - e^{-f} 
 \frac{\partial p}{\partial r}\,  \omega^2 \; , \nonumber \\*
\nonumber \\*
\Gamma_{13} &= & - \, q_{|1}\,\omega^3 = - \,  
e^{-f} \frac{\partial q}{\partial r} \, \omega^3 \; , \nonumber \\*
\Gamma_{23} &= & - \, q_{|2}\,\omega^3 = - \,  
\frac{e^{-p}}{a} \frac{\partial q}{\partial \theta} \, \omega^3\; ,  \nonumber \\*
\Gamma_{14} &= & \, h_{|1}\,\omega^4
= \,  
e^{-f} \frac{\partial h}{\partial r} \, \omega^4 \; , \nonumber \\*
\Gamma_{24} &= & \, h_{|2}\,\omega^4
= \,  
\frac{e^{-p}}{a} \frac{\partial h}{\partial \theta} \, \omega^4  \; ,
\label{gammag}
\end{eqnarray}
with $( \cdot )_{|a} = E_a ( \cdot )$.

The massive spinor of mass $\mu$, minimally coupled to gravity, is described by the $SO(3,1)$ gauge-covariant Dirac
equation
\begin{equation}
\gamma^a \, E_a \Psi + \frac{1}{4} \, \Gamma_{bca} \, \gamma^a \gamma^b  \gamma^c \Psi  + \mu \Psi  = 0 \, .
\label{diraceq}
\end{equation}
In view of the relations (\ref{gammag}), the term expressing the Ricci spin-connection being 
\begin{equation}
 \frac{1}{4} \, \Gamma_{bca} \, \gamma^a \gamma^b \gamma^c = \frac{e^{-f}}{2} \,  \gamma^1  \left[ p+q+h \right],_r +
 \frac{e^{-p}}{2a} \,  \gamma^2 \left[ f+q+h \right],_{\theta} \, ,
 \label{riccispin}
 \end{equation}
the explicit form of the Dirac equation (\ref{diraceq}) reads
\begin{eqnarray}
& &
e^{-f} \gamma^1 \left[ \Psi,_r + \frac{1}{2}
 \left( p+q+h \right),_r  \Psi \right] + \, 
 \frac{e^{-p}}{a} \,  \gamma^2 \left[ \Psi,_{\theta} + \frac{1}{2} \left( f+q+h \right),_{\theta}  \Psi
\right] \nonumber\\&& + \frac{e^{-q}}{b} \gamma^3 \frac{\partial \Psi}{\partial \varphi} + e^{-h} \gamma^4 \, \frac{\partial \Psi}{\partial t}
+ \, \mu  \Psi = 0 \; . \label{dirac1}
\end{eqnarray}

For the concrete example to be tackled in this paper, we shall consider the following metric functions:
\[
e^f = \frac{1}{\Omega \sqrt{R}} \; , \; \;
a e^p = \frac{r}{\Omega \sqrt{P}} \; , \; \;
b e^q = \frac{\sqrt{P} r \sin \theta}{\Omega} \; , \; \;
e^h = \frac{\sqrt{R}}{\Omega} \; ,
\]
where $\Omega = \Omega ( r , \theta)$, $R = R(r)$ and $P = P ( \theta)$. The metric (\ref{generalm}) and and the Ricci-spin connection term (\ref{riccispin}) being now
\begin{equation}
ds^2= \frac{1}{\Omega^2} \left[ \frac{dr^2}{R} - R \, dt^2 + \frac{r^2 d \theta^2}{P} + P r^2 \sin^2 \theta \, d \varphi^2 \right] 
\label{metric2}
\end{equation}
and  
\[
 \frac{1}{4} \, \Gamma_{bca} \, \gamma^a \gamma^b \gamma^c =
\Omega \sqrt{R} \gamma^1 \left[  \frac{1}{r} + \frac{R^{\prime}}{4R} - \frac{3}{2} \frac{\Omega^{\prime}}{\Omega}  \right]  
+ \frac{\Omega \sqrt{P}}{r} \gamma^2 \left[ \frac{\cot \theta}{2} + \frac{\dot{P}}{4P} - \frac{3}{2} \frac{\dot{\Omega}}{\Omega} \right] ,
\]
where $( \, )^{\prime}$ and $\dot{( \, )}$ mean the derivatives with respect to $r$ and $\theta$,
the Dirac equation (\ref{dirac1}) becomes  
\begin{eqnarray}
& &
r \sqrt{R} \gamma^1 \left[ \Psi^{\prime} + \left( \frac{1}{r} + \frac{R^{\prime}}{4R} - \frac{3}{2} \frac{\Omega^{\prime}}{\Omega} \right) \Psi \right] 
+ \sqrt{P} \gamma^2 \left[ \dot{\Psi} + \left( \frac{\cot \theta}{2} + \frac{\dot{P}}{4P} - \frac{3}{2} \frac{\dot{\Omega}}{\Omega} \right) \Psi \right] \nonumber \\*
& & + \, \frac{1}{\sin \theta \sqrt{P}} \, \gamma^3 \Psi,_{\varphi} + \, \frac{r}{\sqrt{R}} \gamma^4 \Psi,_t + \frac{\mu r}{\Omega} \Psi = 0 \, .
\nonumber
\end{eqnarray}

In the particular case of a massless fermion, for which the above equation turns into the form
\begin{eqnarray}
& &
r \sqrt{R} \gamma^1 \left[ \Psi^{\prime} + \left( \frac{1}{r} + \frac{R^{\prime}}{4R} - \frac{3}{2} \frac{\Omega^{\prime}}{\Omega} \right) \Psi \right] 
+ \sqrt{P} \gamma^2 \left[ \dot{\Psi} + \left( \frac{\cot \theta}{2} + \frac{\dot{P}}{4P} - \frac{3}{2} \frac{\dot{\Omega}}{\Omega} \right) \Psi \right] \nonumber \\*
& & + \, \frac{1}{\sin \theta \sqrt{P}} \, \gamma^3 \Psi,_{\varphi} + \, \frac{r}{\sqrt{R}} \gamma^4 \Psi,_t  = 0 \, ,
\label{masslessd}
\end{eqnarray}
with the variables separation of the form:
\begin{equation}
\Psi = \psi ( r , \theta ) e^{i(m \varphi  - \omega t)}   \; ,
\end{equation}
one has to define the function $\psi ( r , \theta)$ as
\begin{equation}
\psi ( r , \theta ) = \Omega^{3/2} R^{-1/4} P^{-1/4}  \, \frac{F(r , \theta)}{r} \, ,
\end{equation}
in order to eliminate the terms containing
the derivatives of the functions $R$, $P$ and $\Omega$.

For the bi-spinor written in terms of two components spinors as
\begin{equation}
F( r , \theta) = \left[ \begin{array}{c}
\zeta  ( r , \theta ) \\ \chi ( r , \theta )
\end{array} \right] ,
\end{equation}
in the Weyl's representation for the $\gamma^i$ matrices,
\begin{equation}
\gamma^1 = -i \beta \, \alpha^3 \; , \; \; \gamma^2  = - i \beta \alpha^1 \; , \; \;
\gamma^3 = -i \beta \alpha^2 \; , \; \; \gamma^4
= - i \beta
\, ,
\end{equation}
with
\[
\alpha^{\mu} = \left(
\begin{array}{cc}
\sigma^{\mu} & 0 \\
0 & - \sigma^{\mu} 
\end{array}
\right)  \; , \; \; \beta = \left(
\begin{array}{cc}
0 &  - {\rm I} \\
{\rm -I} & 0  
\end{array}
\right)  \; , \; \; {\rm so \; that}  \; 
\gamma^5 = \left(
\begin{array}{cc}
 {\rm I} &  0 \\
0 & {\rm -I}   
\end{array}
\right)  ,
\]
where $\sigma^{\mu}$ denotes the usual Pauli matrices,
the equation (\ref{masslessd}) leads to the following system of decoupled equations for the spinors $\zeta$ and $\chi$:
\begin{eqnarray}
& &
r \sqrt{R} \left[ \zeta_1^{\prime} - \frac{i \omega}{R} \zeta_1 \right] 
+ \sqrt{P} \left[ \dot{\zeta_2} + \left( \frac{\cot \theta}{2} + \frac{m}{P \sin \theta}  \right) \zeta_2 \right] = 0 \nonumber \\*
& &
r \sqrt{R} \left[ \zeta_2^{\prime} + \frac{i \omega}{R} \zeta_2 \right] 
- \sqrt{P} \left[ \dot{\zeta_1} + \left( \frac{\cot \theta}{2} - \frac{m}{P \sin \theta}  \right) \zeta_1 \right] = 0 \nonumber \\*
& &
r \sqrt{R} \left[ \chi_1^{\prime} + \frac{i \omega}{R} \chi_1 \right] 
+ \sqrt{P} \left[ \dot{\chi_2} + \left( \frac{\cot \theta}{2} + \frac{m}{P \sin \theta}  \right) \chi_2 \right] = 0 \nonumber \\*
& &
r \sqrt{R} \left[ \chi_2^{\prime} - \frac{i \omega}{R} \chi_2 \right] 
- \sqrt{P} \left[ \dot{\chi_1} + \left( \frac{\cot \theta}{2} - \frac{m}{P \sin \theta}  \right) \chi_1 \right] = 0 \; .
\label{dec}
\end{eqnarray}

Using the standard procedure based on variables separation,
\begin{equation}
\zeta_B = S_B^+ (r) T_B^+ ( \theta ) \; , \;
\chi_B = S_B^- (r) T_B^- ( \theta ) \; , 
\end{equation}
with $B=1,2$, so that $\zeta_1 = S_1^+ T_1^+$, $\zeta_2 = S_2^+ T_2^+$, $\chi_1 = S_1^- T_1^-$, $\chi_2 = S_2^- T_2^-$, one may write down from the system (\ref{dec}) the essential relations:
\begin{eqnarray}
& &
r \sqrt{R} \left[ \frac{d \;}{dr}  - \frac{i \omega}{R} \right] S_1^+ = \lambda S_2^+ \; ;
\nonumber \\*
& &
r \sqrt{R} \left[ \frac{d \;}{dr}  + \frac{i \omega}{R} \right] S_2^+ = \lambda S_1^+
\label{dec1}
\end{eqnarray}
and
\begin{eqnarray}
& &
 \sqrt{P} \left[ \frac{d \; }{d \theta} + \frac{\cot \theta}{2} + \frac{m}{P \sin \theta} \right] T_2^+ = - \lambda T_1^+  \; ; \nonumber \\*
& &
 \sqrt{P} \left[ \frac{d \; }{d \theta} + \frac{\cot \theta}{2} - \frac{m}{P \sin \theta} \right] T_1^+ =  \lambda T_2^+ \; .
 \label{dec2}
\end{eqnarray}
In what it concerns the components $S_B^-$ and $T_B^-$, these are related to $S_B^+$ and $T_B^+$ as: $S_1^- = S_2^+$, $S_2^- = S_1^+$, $T_1^- = T_1^+$ and $T_2^- = T_2^+$. 

Thus, from (\ref{dec1}),
one gets the following system of decoupled equations for the radial functions $S_B^+ = \left \lbrace S_1^+ \, , \, S_2^+ \right \rbrace$
\begin{eqnarray}
& &
\frac{d^2 S_1^+}{dr^2} + \left[ \frac{1}{r} + \frac{R^{\prime}}{2R} \right] \frac{dS_1^+}{dr} + \frac{1}{R} \left[
\frac{\omega^2}{R} - i \omega \left( \frac{1}{r} - \frac{R^{\prime}}{2R} \right) - \frac{\lambda^2}{r^2} \right] S_1^+ = 0 \nonumber \\*
& &
\frac{d^2 S_2^+}{dr^2} + \left[ \frac{1}{r} + \frac{R^{\prime}}{2R} \right] \frac{dS_2^+}{dr} + \frac{1}{R} \left[
\frac{\omega^2}{R} + i \omega \left( \frac{1}{r} - \frac{R^{\prime}}{2R} \right) - \frac{\lambda^2}{r^2} \right] S_2^+ = 0 \; .
\label{dec1f}
\end{eqnarray}

Using the same procedure, the angular equations coming from (\ref{dec2}) can be written in the compact form 
\begin{eqnarray}
& &
\frac{d^2 T_B}{d \theta^2} + \left[ \frac{\dot{P}}{2P} 
 + \cot \theta \right]  \frac{d T_B}{d \theta}
 \nonumber \\*
& &
+ \left[ \frac{\dot{P}}{2P} \left( \frac{\cot \theta}{2} \pm \frac{m}{P \sin \theta} \right) 
- \left( \frac{\cot \theta}{2} \mp \frac{m}{P \sin \theta} \right)^2 - \frac{\lambda^2}{P} - \frac{1}{2} \right] T_B = 0 \, .
\label{dec2f}
\end{eqnarray}
where $B=1,2$ and the upper signs correspond to $B=1$ while the lower signs to $B=2$.

\section{Accelerating charged black holes}

In the followings, we are going to employ the method developed in the previous section and consider in the general metric (\ref{metric2}) the nonlinear superposition of the Reissner--Nordstrom
black hole solution and the Rindler flat spacetime associated with uniformly
accelerated observers, i.e. the charged C-metric \cite{grifpod}:
\begin{eqnarray}
& & R_{RNA}(r) = \left( 1 - \frac{2M}{r} + \frac{Q^2}{r^2} \right) \left( 1-A^2 r^2 \right) \; , 
\nonumber \\* 
& &
P( \theta ) = 1-2MA \cos \theta +A^2 Q^2 \cos^2 \theta \; ,
\nonumber \\*
& &
\Omega ( r , \theta ) = 1 - Ar \cos \theta \, .
\label{chargedC}
\end{eqnarray}
In (\ref{chargedC}), $M$, $Q$ and $A$ are the mass, the charge and the acceleration, and the vector potential for the electromagnetic field is given by
\[
{\rm A} = - \frac{Q}{r} \, dt \, .
\]
There is a curvature singularity at $r=0$, the acceleration horizon at $r=1/A$ and the  inner
and outer black hole horizons
\begin{equation}
r_{\pm} = M \pm \varepsilon \; , \; \varepsilon = \sqrt{M^2 - Q^2} \,  .
\end{equation}

For $R_{RNA}$ written in the factorized form \cite{Hong:2003gx}
\[
R_{RNA} = \left(1- \frac{r_+}{r} \right) \left(1-\frac{r_-}{r} \right) \left(1-A^2 r^2 \right) \, ,
\]
it is not a difficult task to find, using Maple, the solutions of the corresponding radial equation coming from (\ref{dec1f}). 
With the notation $\eta = 2 A \varepsilon$ and up to some normalization constants, these are expressed in terms of Heun general functions \cite{Ronveaux}, \cite{Slavyanov} as:
\begin{eqnarray}
S_1^+ & = & C_1 \,  {\cal F} (r \, )HeunG  \left[ a, \, q , \, \alpha , \beta , \, \gamma , \, \delta , \, y
\right]  \nonumber \\*
 & & + C_2 \,  {\cal F} (r) \, y^{1- \gamma} HeunG  \left[ a, \, q^{\prime} , \, \alpha^{\prime} , \beta^{\prime} , \, \gamma^{\prime} , \, \delta , \, y
\right] 
\end{eqnarray}
where
\begin{eqnarray}
{\cal F} (r) & = & \frac{\sqrt{r^2 \, R_{RNA}}}{(1-Ar)^2} \,  (r-r_+)^{- \frac{i \omega A r_+^2}{\eta \left( 1-A^2 r_+^2 \right)} }  \,  (r-r_-)^{\frac{i \omega A r_-^2}{\eta \left( 1-A^2 r_-^2 \right)} } \nonumber \\*
& \times &  \,  (1+Ar)^{- \frac{i \omega}{2A (1+Ar_+)(1+Ar_-)}}  \,  (1-Ar)^{\frac{i \omega}{2A (1-Ar_+)(1-Ar_-)}}   \, ,
\label{fr}
\end{eqnarray}
the variable of the Heun functions is
\[
y = \frac{A(1- A r_+)}{\eta} \left[ \frac{r-r_-}{1-Ar} \right] ,
\]
while the two set of parameters are given by:
\begin{eqnarray}
& &
a = - \, \frac{(1-Ar_+)(1+Ar_-)}{2 \eta}  \; , \; q = \frac{3}{2} + \frac{1}{2 \eta} \left[ A^2 Q^2 + \lambda^2 - 1 + \frac{4i \omega r_-}{1-Ar_-} \right] \; ,
\nonumber \\*
& &
\alpha = 2 \; ,  \; \beta = \frac{3}{2} - \frac{i \omega}{A(1-Ar_+)(1-Ar_-)} \; ,
\nonumber \\*
& & \gamma = \frac{3}{2} + \frac{2i \omega Ar_-^2}{\eta(1-A^2 r_-^2)}\; ,  \;  \delta =\frac{3}{2} - \frac{2i \omega Ar_+^2}{\eta(1-A^2 r_+^2)}  
\end{eqnarray}
and
\begin{equation}
q^{\prime} = q + ( a \delta +  \alpha + \beta - \gamma - \delta  +1)(1 - \gamma)  , \;
\alpha^{\prime} = \alpha + 1 - \gamma  , \;
\beta^{\prime} = \beta + 1 - \gamma  ,
\gamma^{\prime} = 2 - \gamma \; .
\label{param1}
\end{equation}

The Heun's general equation has 4 regular singularities situated at $y=0$, $y=1$, $y=a$ and $y = \infty$ \cite{Ronveaux}, \cite{Slavyanov} meaning, in our case, $r = r_{\pm}$ and $r = \pm 1/A$. The six free parameters play different roles. Thus, $a$ is the singularity parameter, $q$ is the accessory parameter, while the others are exponent parameters.

One may look for a polynomial form of the Heun functions in (\ref{fr}), by imposing that the third parameter is satisfying the necessary (not sufficient) condition $\alpha = - n$, with $n$ a positive integer and $q$ has one of a finite number of characteristic values, which fixes the value of $\lambda$.
This can be done for $\alpha^{\prime}$ in (\ref{param1}), leading to the quantized imaginary spectrum
\begin{equation}
\omega = - i \left( n + \frac{3}{2} \right) \frac{\varepsilon(1-A^2r_-^2)}{r_-^2} \; .
\end{equation}

Finally, for $P( \theta )$ given in (\ref{chargedC}), written as
\[
P( \theta ) = \left( 1 - A r_+ \cos \theta \right) \left( 1 - A r_- \cos \theta \right) ,
\] 
the solutions to the angular equations coming from (\ref{dec2f}) has a very similar form, namely
\begin{eqnarray}
T_1 & = & C_1 \,    {\cal G} ( \theta  ) \, HeunG  \left[ a, \, q , \, \alpha , \beta , \, \gamma , \, \delta , \, \tau
\right]  \nonumber \\*
 & & + \, C_2 \, {\cal G} ( \theta  ) \, \tau^{1- \gamma} HeunG  \left[ a, \, q^{\prime} , \, \alpha^{\prime} , \beta^{\prime} , \, \gamma^{\prime} , \, \delta , \, \tau
\right] 
\end{eqnarray}
and $m \to -m$ for $T_2$, where
\begin{eqnarray}
{\cal G} ( \theta  )  & = & \frac{\sqrt{  P( \theta ) \, \sin \theta}}{(1-Ar_+ \cos \theta )^2} \,  \left( 1 - A r_+ \cos \theta \right)^{\frac{m A^2 r_+^2}{\eta \left( 1-A^2 r_+^2 \right)} }  \,  \left( 1 - A r_- \cos \theta \right)^{- \frac{m A^2 r_-^2}{\eta \left( 1-A^2 r_-^2 \right)} } \nonumber \\*
& \times &  \,  \left( \sin \frac{\theta}{2} \right)^{- \frac{m}{(1-Ar_+)(1-Ar_-)}}  \,  \left( \cos \frac{\theta}{2} \right)^{\frac{m}{(1+Ar_+)(1+Ar_-)}}   \, ,
\label{gtheta}
\end{eqnarray}
the variable of the Heun functions being:
\[
\tau = \left( 1 - A r_+ \right) \frac{\cos^2 \frac{\theta}{2}}{1- A r_+ \cos \theta}  ,
\]
and the two set of parameters are:
\begin{eqnarray}
& &
a = - \, \frac{(1-Ar_+)(1+Ar_-)}{2 \eta}  \; , \; q = \frac{3}{2} + \frac{1}{2 \eta} \left[ A^2 Q^2 - \lambda^2 - 1 + \frac{4 mAr_+}{1+Ar_+} \right] \; ,
\nonumber \\*
& &
\alpha = 2 \; ,  \; \beta = 
\frac{3}{2} - \frac{2 m A^2 r_+^2}{\eta(1-A^2 r_+^2)}  \; ,
\nonumber \\*
& &
\gamma = \frac{3}{2} + \frac{m}{(1+Ar_+)(1+Ar_-)} \; , \; 
 \delta = \frac{3}{2} - \frac{m}{(1-Ar_+)(1-Ar_-)} \; ,
\end{eqnarray}
and $q^{\prime}$, $\alpha^{\prime}$, $\beta^{\prime}$, $\gamma^{\prime}$ satisfying the same relations as in (\ref{param1}).

At the end of this section, let us mention briefly the case of the extreme Reissner-Nordstrom accelerated black hole,
with $Q= \pm M$, for which the radial function in (\ref{chargedC}) is 
\begin{equation}
R (r) = \left( 1 - \frac{M}{r} \right)^2 \left( 1- A^2 r^2 \right) .
\label{cextremal}
\end{equation}

The solutions to the radial equations (\ref{dec1f})  are expressed in terms of Heun confluent functions \cite{Ronveaux}, \cite{Slavyanov} as:
\begin{eqnarray}
S_1^+ & =  & \left( 1-Ar \right)^{\frac{\gamma}{2}+ \frac{1}{4}} e^{\frac{\alpha z}{2}}  \times
\nonumber \\*
& & 
 \left \lbrace C_1 
\left( 1+Ar \right)^{- \frac{\beta}{2}+ \frac{1}{4}} (r-M)^{- \frac{1}{2} - \frac{i \omega (1+ \sigma^2)}{A(1- \sigma^2)^2} } 
 HeunC  \left[ \alpha ,  - \beta , \, \gamma , \, \delta , \, , \, \eta , \, z
\right]  \right. \nonumber \\*
& & \left.  + C_2 \, \left( 1+Ar \right)^{\frac{\beta}{2}+ \frac{1}{4}} (r-M)^{- 1 - \frac{2 i \omega \sigma}{A(1- \sigma^2)^2}} 
HeunC  \left[ \alpha ,   \beta , \, \gamma , \, \delta , \, , \, \eta , \, z
\right]  \right \rbrace
\end{eqnarray}
where $\sigma = MA$,
while the variable $z$ and the parameters of the Heun functions are respectively given by:
\[
z = \frac{1- \sigma}{2A} \left[ \frac{1+Ar}{r-M} \right] ,
\]
and 
\begin{eqnarray}
& &
\alpha = \frac{4i \omega \sigma^2}{A(1-\sigma^2)^2} \; ,  \; \beta = \frac{1}{2} - \frac{i \omega}{A(1+ \sigma)^2} \; ,
\nonumber \\* & &
\gamma =  \frac{1}{2} + \frac{i \omega}{A(1- \sigma)^2} \;  ,   \;
 \delta =\frac{\alpha}{2} \left( 1-4i \omega M \right) , 
\nonumber \\*
& &\eta = 
\frac{i \omega}{A (1+ \sigma)(1- \sigma^2)} \left[ \sigma + \frac{i \omega (1+3 \sigma)}{2A ( 1+ \sigma)^2} \right] + \frac{3}{8} - \frac{\lambda^2}{1- \sigma^2} \; .
\end{eqnarray}

One may easily check that the orbital equations, with
\[
P( \theta ) = \left( 1- MA \cos \theta \right)^2 \; ,
\]
are also satisfied by Heun confluent functions.

\section{Particular Cases}

\subsection{The vacuum C-metric in spherical-type coordinates}

The Dirac equation for massless particles evolving in the vacuum C-metric spacetime has been exactly solved in \cite{Bini:2015xpa}. By defining the Kinnersley-like null frame and using the NP formalism, the Dirac equation
has been separated into one-dimensional radial and angular parts, both being satisfied by Heun general functions.

In the followings, we consider in the general metric (\ref{metric2}) the nonlinear superposition of the Schwarzschild
black hole solution and the Rindler flat spacetime, i.e.  
\begin{eqnarray}
& & R_c(r) = \left( 1 - \frac{2M}{r} \right) \left( 1-A^2 r^2 \right) \; , \;
P( \theta ) = 1-2MA \cos \theta \; ,
\nonumber \\*
& &
\Omega ( r , \theta ) = 1 - Ar \cos \theta \, .
\label{cmetric}
\end{eqnarray}

The form (\ref{cmetric}) has the important property that it reduces to the
spherically symmetric Schwarzschild solution, when the acceleration parameter vanishes $A=0$.
Thus, the positive constant $M$ is the mass of the
source, while $r$ is the Schwarzschild-like radial coordinate, with a black hole horizon at $r=2M$.

One can easily check using the Maple Soft \cite{Maple}, that the radial equations coming from (\ref{dec1f}), with $R_c(r)$ given in (\ref{cmetric}), i.e.
\begin{eqnarray}
& &
\frac{d^2 S_1^+}{dr^2} + \frac{r(1-2A^2r^2)-M(1-3A^2r^2)}{r^2 \, R_c} \frac{dS_1^+}{dr} 
\nonumber \\* & &
+ \frac{1}{R_c} \left[ \frac{\omega^2}{R_c}
-  i \omega \, \frac{r-M(3-A^2r^2)}{r^2 \, R_c} - \frac{\lambda^2}{r^2} \right] S_1^+  = 0 \, ,
\end{eqnarray}
and similarly for $S_2^+$, where $\omega \to - \omega$,
has the solutions expressed in terms of Heun general functions as
\begin{eqnarray}
& & S_1^+ \, = \, ( 1+Ar)^{\frac{i \omega M}{\eta ( 1+ \eta)}}  ( 1-Ar)^{\frac{-i \omega M}{\eta ( 1- \eta)}} 
( r-2M)^{\frac{2i \omega M}{1-\eta^2}}  \nonumber \\*
&  & \times \left \lbrace C_1 HeunG  \left[ a, \, q^{\prime} , \, \alpha^{\prime} , \beta^{\prime} , \, \gamma^{\prime} , \, \delta , \, y
\right] + C_2 \sqrt{y} HeunG  \left[ a, \, q^{\prime \prime} , \, \alpha^{\prime \prime} , \beta^{\prime \prime} , \, \gamma^{\prime \prime} , \, \delta , \, y
\right] \right \rbrace , \label{S1plus}
\end{eqnarray}
with $\eta = 2MA$, the variable
\[
y = \frac{(1- \eta) r}{2M(1-Ar)} \, ,
\]
and the parameters
\begin{eqnarray}
& &
a = \frac{1}{2} \left( 1 - \frac{1}{\eta} \right) \; , \; q^{\prime} = \frac{\lambda^2}{2 \eta} \; , \; q^{\prime \prime} = q^{\prime} + \frac{3}{8} - \frac{1}{8 \eta} \; ,
\nonumber \\*
& &
\alpha^{\prime} = 0 \; , \; \alpha^{\prime \prime} = \frac{1}{2} \; , \; \beta^{\prime} = \frac{1}{2} + \frac{2i \omega M}{\eta(1- \eta )} \; , \;
\beta^{\prime \prime} = \beta^{\prime} + \frac{1}{2} \; ,
\nonumber \\*
& & \gamma^{\prime} = \frac{1}{2} \; , \; \gamma^{\prime \prime} = \frac{3}{2}  \; , \; \delta = \frac{1}{2} + \frac{4i \omega M}{1-\eta^2}  \, .
\label{param2}
\end{eqnarray}

The general Heun equation in its canonical form given in literature \cite{Ronveaux}, \cite{Slavyanov} has regular singularities at $y=0$, $y=1$, $y=a$ and $y = \infty$, meaning, in our case, $r=0$, $r=2M$ and $r = \pm 1/A$,
with the expansion around $y=0$ given by
\[
HeunG \approx 1 + \frac{q}{\gamma a} y + {\cal O} (y^2 ) \, .
\]

In the particular case $\omega =0$, so that the parameters (\ref{param2}) get real values, the procedure of reducing the Heun general functions to Heun polynomials is discussed in \cite{Kar:2017pzt}. Thus,
for the case $\alpha^{\prime} =0$ and $\gamma^{\prime} = 1/2$, the first Heun function in (\ref{S1plus}) turns into a constant, while for  $\alpha^{\prime \prime} =1/2$ and $\gamma^{\prime \prime} = 3/2$, the second Heun function is $1/ \sqrt{y}$.

As for the angular equations (\ref{dec2f}), these are also satisfied by Heun general functions which can be obtained from (\ref{gtheta}), for $r_+ = 2M$, $r_- =0$ and $\eta = 2MA$.

\subsection{The Reissner--Nordstrom geometry}

As another physically important example, let us consider the case where the acceleration parameter in the charged C-metric is zero $A=0$, meaning that $P= \Omega =1$ in \ref{metric2}. Thus, for the function
\begin{equation}
R_{RN} = 1 - \frac{2GM}{c^2r} + \frac{GQ^2}{c^4r^2} \, ,
\end{equation}
that in natural units $G=c=1$ turns into the well-known Reissner--Nordstrom expression
\begin{equation}
R_{RN} = 1 - \frac{2M}{r} + \frac{Q^2}{r^2} \, ,
\label{RN}
\end{equation}
the radial equations (\ref{dec1f}) become
\begin{eqnarray}
\frac{d^2 S_B^+}{dr^2} + \frac{r-M}{r^2 \, R_{RN}} \frac{dS_B^+}{dr} 
+ \left[ \frac{\omega^2}{R_{RN}^2}
\mp \frac{ i \omega(r^2-3Mr+2 Q^2)}{r^3 \, R_{RN}^2} - \frac{\lambda^2}{r^2 \, R_{RN}} \right] S_B^+  = 0 \, ,
\nonumber \\*
\end{eqnarray}
where $-i \omega$ stands for $B=1$, while $+i \omega$ is for $B=2$.

The solutions are expressed in terms of Heun confluent functions as
\begin{equation}
S_1^+ = e^{i \omega r}\,  r^{\frac{1}{2} \pm \beta} \, R_{RN}^{\frac{1}{4} \pm \frac{\beta}{2} } \, HeunC [ \alpha , \, \pm \beta , \, \gamma , \, \delta , \, \eta , \, y ] ,
\label{S1plusRN}
\end{equation}
of variable
\[
y = \frac{r-M+\varepsilon}{2 \varepsilon} = \frac{r - r_-}{r_+ - r_-} \, ,
\]
and parameters
\begin{eqnarray}
& &
\alpha = 4i \omega \varepsilon \, , \; 
\beta = - \frac{1}{2} - \frac{i \omega r_-^2}{\varepsilon}  \;  , \;
\gamma = \frac{1}{2} - \frac{i \omega r_+^2}{\varepsilon}
\; , \; 
\delta = 2 \omega \varepsilon (4M \omega -i ) \; , \nonumber \\*
& &
\eta = \omega  (4M \omega -i ) ( M - \varepsilon) - \frac{\omega^2(M^4 - \varepsilon^4)}{\varepsilon^2} + \frac{3}{8} -
\lambda^2 \; .
\label{param3}
\end{eqnarray}
As in the previous section,  $r_{\pm} = M \pm \varepsilon$ are the outer and inner horizons, with $\varepsilon = \sqrt{M^2 - Q^2}$ and the relation (\ref{RN}) can be written as
\[
R_{RN} = \left( 1 - \frac{r_-}{r} \right) \left( 1 - \frac{r_+}{r} \right) 
\]
pointing out the two regular singular points $r_{\pm}$ which are to be treated on equal footing.
 
Since $\beta$ is a non-integer quantity, the two functions in (\ref{S1plusRN}) form linearly independent solutions of the confluent Heun differential equation.

For the  polynomial form of the confluent Heun functions, one has to impose the necessary condition \cite{Ronveaux}, \cite{Slavyanov}
\begin{equation}
\frac{\delta}{\alpha} = - \left[ n+1 + \frac{\beta+\gamma}{2} \right] \, .
\label{cond1}
\end{equation}
In view of the parameters in (\ref{param3}), it turns out that only the component multiplied by $C_1$ gets a polynomial expression, the imaginary quasispectrum being given by the quantization law
\begin{equation}
\omega =  - \, i \left( n+ \frac{1}{2} \right) \frac{\varepsilon}{r_+^2} .
\label{cond2}
\end{equation}

If one further sets the charge parameter to zero $Q=0$, note that the particular case of the Dirac equation describing massless fermions in Schwarzschild spacetime has been investigated recently, using both  the NP formalism \cite{Al-Badawi:2017fja} and the partial wave analysis \cite{Cotaescu:2016aty}. 

However, this problem can be analytically solved within the approach developed in the present work, leading to confluent Heun functions. Indeed, with 
\[
R_S(r) = 1 - \frac{2M}{r} \, ,
\]
the corresponding radial equations (\ref{dec1f}), i.e.
\begin{eqnarray}
\frac{d^2 S_B^+}{dr^2} + \frac{r-M}{r^2 \, R_S} \frac{dS_B^+}{dr} 
+ \left[ \frac{\omega^2}{R_S^2}
\mp \frac{ i \omega(r-3M)}{r^2 \, R_S^2} - \frac{\lambda^2}{r^2 \, R_S} \right] S_B^+  = 0 \, ,
\end{eqnarray}
have the solution
\begin{eqnarray}
S_1^+ & = & C_1 \,  e^{i \omega r} (r-2M)^{2i \omega M} HeunC \left[ \alpha , \, \beta , \, \gamma , \, \delta , \, \eta , \, \frac{r}{2M} \right] 
\nonumber \\* & &
+ \, C_2 \, \sqrt{r} e^{i \omega r} (r-2M)^{2i \omega M} HeunC \left[ \alpha , \, - \beta , \, \gamma , \, \delta , \, \eta , \, \frac{r}{2M} 
\right] 
\label{S1plusSchw}
\end{eqnarray}
and its complex conjugate, with the parameters of the confluent Heun functions being given by
\begin{equation}
\alpha = 4i \omega M \; , \; \beta = - \frac{1}{2} \, , \, \gamma = 4i \omega M - \frac{1}{2} \, , \,
\delta = 2 \omega M ( 4 \omega M -i) \, , \,
\eta = \frac{3}{8} - \lambda^2 \, .
\end{equation}

The solutions to Heun's confluent equations are computed as power series expansions around the regular singular point $x= r/(2M)=0$. The series converges for $x <1$, i.e. $r < 2M$, where the second regular singularity is located. An analytic continuation of the HeunC function is obtained by expanding the solution around the regular singularity $x=1$, and overlapping the series. 

For the asymptotic behavior at infinity, one may use the formula \cite{Ronveaux}
\[
HeunC \left[ \alpha , \, \beta , \, \gamma , \, \delta , \, \eta , \, x \right]
\approx D_1 x^{- \left[ \frac{\beta+\gamma+2}{2} + \frac{\delta}{\alpha} \right]}
+ D_2 e^{- \alpha x} x^{- \left[ \frac{\beta+\gamma+2}{2} - \frac{\delta}{\alpha} \right]}
\]
and the expression (\ref{S1plusSchw}) turns into the simplified form
\begin{eqnarray}
& &
S_1^+ \approx \frac{1}{\sqrt{r}} \left[ D_1 e^{i \omega r} r^{2i \omega M + \frac{1}{2}} +
D_2 e^{-i \omega r} r^{-2i \omega M - \frac{1}{2}} \right] \nonumber \\*
& & \approx \frac{D}{\sqrt{r}}  \sin \left[ \omega r + \left( 2 \omega M - \frac{i}{2} \right) \log r + \phi ( \omega ) \right] ,
\label{asympt}
\end{eqnarray}
where $\phi ( \omega )$ is the phase shift and $S_2^+ = S_1^+ ( \omega \to - \omega )$.

As for the angular parts $T_B^{\pm}$, one has to go back to the equations (\ref{dec2f}) which, for $P=1$, take the form 
\begin{equation}
\frac{d^2T_B^+}{d \theta^2} + \cot \theta \frac{dT_B^+}{d \theta} - \left[ \frac{( \cos \theta \mp 2m)^2}{4 \sin^2 \theta} - \lambda^2 + \frac{1}{2} \right] T_B^+ = 0  \, ,
\end{equation}
with the solutions expressed in terms of hypergeometric functions $F_{12}$ as
\begin{eqnarray}
& &
T_B^+ \, = \, \frac{C_1}{\sqrt{\sin \theta}} \left( \tan \frac{\theta}{2} \right)^{\pm m} F_{12} \left[ - \lambda , \, \lambda , \, \pm m + \frac{1}{2} , \, \sin^2 \frac{\theta}{2} \right] \nonumber \\*
& & + \,
C_2 \, \sqrt{\tan \frac{\theta}{2}}\,  ( \sin \theta ) ^{\mp m} \,  F_{12} \left[ \frac{1}{2} - \lambda \mp m , \, \frac{1}{2} + \lambda \, \mp m , \, \frac{3}{2} \mp \lambda , \, \sin^2 \frac{\theta}{2} \right]  ,  \; \; \; \;   \; \;
\end{eqnarray}
and the same functions for $T_B^-$.

\subsection{The BBMB metric}

As a final example, let us consider the metric corresponding to:
\begin{equation}
R (r) = \left( 1 - \frac{M}{r} \right)^2 \, ,
\label{BBMB}
\end{equation}
which can be obtained from (\ref{cextremal}), for $A=0$.

This metric, which can be expressed as an exact solution to the Einstein conformal scalar equations is called the BBMB solution \cite{Bocharova:1970skc}, \cite{Bekenstein:1974sf}
and one may notice that the spacetime is the same with the extremal Reissner-Nordstrom black hole, for $Q= \pm M$.
In (\ref{BBMB}), $M$ is the mass of black hole and the unique event horizon is located at $r=M$, where the scalar field diverges.
 
In the general equations (\ref{dec1f}) which get the form
\begin{eqnarray}
\frac{d^2 S_B^+}{dr^2} + \frac{1}{r-M}\frac{dS_B^+}{dr} 
+ \frac{1}{(r-M)^2} \left[ \frac{\omega^2 r^4}{(r-M)^2}
\mp \frac{ i \omega r(r-2M)}{r-M} - \lambda^2 \right] S_B^+  = 0 \, ,
\end{eqnarray}
we introduce the new variable
\[
y = 1 - \frac{2M}{r} \, ,
\]
pointing out the special value $r=2M$, where the photon surface is located \cite{Tomikawa:2017vun}. Thus, for $S_1^+$, we get the new equation
\begin{equation}
(1-y^2)^2 \frac{d^2 S_1^+}{dy^2} -2y(1-y^2) \frac{d S_1^+}{dy}
+ 4 \left[ \frac{16 \omega^2 M^2}{(1-y^2)^2} - \frac{4i \omega My}{1-y^2} - \lambda^2 \right] S_1 = 0, 
\end{equation}
whose solutions are
\begin{equation}
S_1^+ \, = \, C \exp   \left[ \mp \frac{4i \omega My}{1-y^2} \right] HeunD \left[ \pm \alpha , \, \beta , \, \gamma , \, \delta ,  \, y \right]   ,
\end{equation}
with the following parameters of the Heun double confluent functions  
\begin{equation}
\alpha = 8i \omega M , \; \beta = \frac{\alpha^2}{4} - 4 \lambda^2  , \; \gamma = 2 \alpha  , \;
\delta = \frac{3 \alpha^2}{4} + 4 \lambda^2 
\end{equation}
and the conditions at the origin become $HeunD (0) =1$ and $HeunD^{\prime} (0) =0$.

\section{Conclusions}

In the present paper, we presented an alternative approach to the usual method based on the NP formalism, for deriving the $SO(3,1)-$gauge covariant Dirac equation in curved spacetimes. This procedure applies to general static prolate metric of the form (\ref{generalm}). For massless fermions, the general equation (\ref{dirac1}), in its concrete form (\ref{masslessd}) can be separated into its radial and angular parts. These have been solved in the general case of accelerating charged black holes, the solutions being expressed in terms of general Heun functions.

As expected, in the case of the vacuum C-metric, we obtain similar results as the ones derived in \cite{Bini:2015xpa}, where the authors used a Kinnersley-like null frame and the NP formalism.

When the acceleration parameter is zero, one deals with the Reissner--Nordstrom spacetime and the solution to the Dirac equation is given in terms of confluent Heun functions. By imposing the parameters (\ref{param3}) to satisfy the condition (\ref{cond1}), the confluent Heun function gets a polynomial form. Moreover, the imaginary quasispectrum of massless fermions is given by the remarkably simple analytical formula (\ref{cond2}), pointing out
the characteristic timescale $\tau$ determined by the fundamental quasinormal resonant frequency corresponding to $n=0$, i.e.
$\tau = 2r_+^2 / \varepsilon$.

Far from the Reissner--Nordstrom or Schwarzschild black holes, asymptotic forms of the radial functions, as the one in (\ref{asympt}), can be used to
investigate, for example, the scattering of 
astrophysical neutrinos \cite{Vieira:2016ubt}.

Even though there is known by now a rich collection of relations and properties of Heun functions of all kinds \cite{Ishkhanyan1}, \cite{Ishkhanyan2} for a comprehensive description of classes of exact solutions to the Teukolsky Master Equation, which is a basic tool in black hole physics, expressed in terms of the confluent Heun functions, we recommend \cite{Fiziev:2009wn}.

Finally, one can notice that for the metric (\ref{BBMB}), which is the extremal form of the Reissner-Nordstrom solution, the confluent Heun functions (CHF) have turned into double confluent Heun functions, through an additional confluence process, with the two regular singularities of the confluent Heun equation coalescing into one irregular singularity at the origin. The irregular singularities at 0 and $\infty$, have been further relocated at $y= \mp 1$, leaving the origin as a regular point. 

For $y$ in the range $y \in ( -1  , 1)$, meaning $r \in( M ,  \infty )$, there is a maximum value at $y=0$, i.e. $r=2M$.
For $r<M$, which corresponds to $y$ outside the convergence unit circle, the expressions of the Heun double confluent functions can be constructed using the identity
\[
HeunD [ \alpha , \beta , \gamma , \delta , y ] =  HeunD \left[ - \alpha , - \delta, - \gamma, - \beta , \frac{1}{y} \right] .
\]

\begin{flushleft}
\begin{Large}
{\bf Acknowledgement}
\end{Large}
\end{flushleft}
This work was supported by a grant of Ministery of Research and Innovation, CNCS - UEFISCDI, project number PN-III-P4-ID-PCE-2016-0131, within PNCDI III.

\end{document}